\documentclass{amsart}
\usepackage{amsfonts}
\usepackage{amsmath,amscd}
\usepackage{amsthm}
\usepackage{amssymb}
\usepackage{latexsym}
\usepackage{color}

 \usepackage[curve,matrix,arrow,color]{xy}

\newtheorem{notation}{Notation}

\numberwithin{equation}{section}
\begin{document}
\newcommand{\beqa}{\begin{eqnarray}}
\newcommand{\eeqa}{\end{eqnarray}}
\newcommand{\thmref}[1]{Theorem~\ref{#1}}
\newcommand{\secref}[1]{Sect.~\ref{#1}}
\newcommand{\lemref}[1]{Lemma~\ref{#1}}
\newcommand{\propref}[1]{Proposition~\ref{#1}}
\newcommand{\corref}[1]{Corollary~\ref{#1}}
\newcommand{\remref}[1]{Remark~\ref{#1}}
\newcommand{\er}[1]{(\ref{#1})}
\newcommand{\nc}{\newcommand}
\newcommand{\rnc}{\renewcommand}

\nc{\cal}{\mathcal}

\nc{\goth}{\mathfrak}
\rnc{\bold}{\mathbf}
\renewcommand{\frak}{\mathfrak}
\renewcommand{\Bbb}{\mathbb}

\def\cN{{\cal N}}
\newcommand{\hs}[1]{\hspace{#1 mm}}
\newcommand{\mb}[1]{\hs{4}\mbox{#1}\hs{4}}
\newcommand{\id}{\text{id}}
\nc{\Cal}{\mathcal}
\nc{\Xp}[1]{X^+(#1)}
\nc{\Xm}[1]{X^-(#1)}
\nc{\on}{\operatorname}
\nc{\ch}{\mbox{ch}}
\nc{\Z}{{\bold Z}}
\nc{\J}{{\mathcal J}}
\nc{\C}{{\bold C}}
\nc{\Q}{{\bold Q}}

\nc{\bak}{\bar{k}}

\renewcommand{\P}{{\mathcal P}}
\nc{\N}{{\Bbb N}}
\nc\beq{\begin{equation}}
\nc\enq{\end{equation}}
\nc\lan{\langle}
\nc\ran{\rangle}
\nc\bsl{\backslash}
\nc\mto{\mapsto}
\nc\lra{\leftrightarrow}
\nc\hra{\hookrightarrow}
\nc\sm{\smallmatrix}
\nc\esm{\endsmallmatrix}
\nc\sub{\subset}
\nc\ti{\tilde}
\nc\nl{\newline}
\nc\fra{\frac}
\nc\und{\underline}
\nc\ov{\overline}
\nc\ot{\otimes}
\nc\bbq{\bar{\bq}_l}
\nc\bcc{\thickfracwithdelims[]\thickness0}
\nc\ad{\text{\rm ad}}
\nc\Ad{\text{\rm Ad}}
\nc\Hom{\text{\rm Hom}}
\nc\End{\text{\rm End}}
\nc\Ind{\text{\rm Ind}}
\nc\Res{\text{\rm Res}}
\nc\Ker{\text{\rm Ker}}
\rnc\Im{\text{Im}}
\nc\sgn{\text{\rm sgn}}
\nc\tr{\text{\rm tr}}
\nc\Tr{\text{\rm Tr}}
\nc\supp{\text{\rm supp}}
\nc\card{\text{\rm card}}
\nc\bst{{}^\bigstar\!}
\nc\he{\heartsuit}
\nc\clu{\clubsuit}
\nc\spa{\spadesuit}
\nc\di{\diamond}
\nc\cW{\cal W}
\nc\cG{\cal G}
\nc\al{\alpha}
\nc\bet{\beta}
\nc\ga{\gamma}
\nc\de{\delta}
\nc\ep{\epsilon}
\nc\io{\iota}
\nc\om{\omega}
\nc\si{\sigma}
\rnc\th{\theta}
\nc\ka{\kappa}
\nc\la{\lambda}
\nc\ze{\zeta}

\nc\vp{\varpi}
\nc\vt{\vartheta}
\nc\vr{\varrho}

\nc\Ga{\Gamma}
\nc\De{\Delta}
\nc\Om{\Omega}
\nc\Si{\Sigma}
\nc\Th{\Theta}
\nc\La{\Lambda}

\nc\boa{\bold a}
\nc\bob{\bold b}
\nc\boc{\bold c}
\nc\bod{\bold d}
\nc\boe{\bold e}
\nc\bof{\bold f}
\nc\bog{\bold g}
\nc\boh{\bold h}
\nc\boi{\bold i}
\nc\boj{\bold j}
\nc\bok{\bold k}
\nc\bol{\bold l}
\nc\bom{\bold m}
\nc\bon{\bold n}
\nc\boo{\bold o}
\nc\bop{\bold p}
\nc\boq{\bold q}
\nc\bor{\bold r}
\nc\bos{\bold s}
\nc\bou{\bold u}
\nc\bov{\bold v}
\nc\bow{\bold w}
\nc\boz{\bold z}

\nc\ba{\bold A}
\nc\bb{\bold B}
\nc\bc{\bold C}
\nc\bd{\bold D}
\nc\be{\bold E}
\nc\bg{\bold G}
\nc\bh{\bold H}
\nc\bi{\bold I}
\nc\bj{\bold J}
\nc\bk{\bold K}
\nc\bl{\bold L}
\nc\bm{\bold M}
\nc\bn{\bold N}
\nc\bo{\bold O}
\nc\bp{\bold P}
\nc\bq{\bold Q}
\nc\br{\bold R}
\nc\bs{\bold S}
\nc\bt{\bold T}
\nc\bu{\bold U}
\nc\bv{\bold V}
\nc\bw{\bold W}
\nc\bz{\bold Z}
\nc\bx{\bold X}

\nc\ca{\mathcal A}
\nc\cb{\mathcal B}
\nc\cc{\mathcal C}
\nc\cd{\mathcal D}
\nc\ce{\mathcal E}
\nc\cf{\mathcal F}
\nc\cg{\mathcal G}
\rnc\ch{\mathcal H}
\nc\ci{\mathcal I}
\nc\cj{\mathcal J}
\nc\ck{\mathcal K}
\nc\cl{\mathcal L}
\nc\cm{\mathcal M}
\nc\cn{\mathcal N}
\nc\co{\mathcal O}
\nc\cp{\mathcal P}
\nc\cq{\mathcal Q}
\nc\car{\mathcal R}
\nc\cs{\mathcal S}
\nc\ct{\mathcal T}
\nc\cu{\mathcal U}
\nc\cv{\mathcal V}
\nc\cz{\mathcal Z}
\nc\cx{\mathcal X}
\nc\cy{\mathcal Y}

\nc\e[1]{E_{#1}}
\nc\ei[1]{E_{\delta - \alpha_{#1}}}
\nc\esi[1]{E_{s \delta - \alpha_{#1}}}
\nc\eri[1]{E_{r \delta - \alpha_{#1}}}
\nc\ed[2][]{E_{#1 \delta,#2}}
\nc\ekd[1]{E_{k \delta,#1}}
\nc\emd[1]{E_{m \delta,#1}}
\nc\erd[1]{E_{r \delta,#1}}

\nc\ef[1]{F_{#1}}
\nc\efi[1]{F_{\delta - \alpha_{#1}}}
\nc\efsi[1]{F_{s \delta - \alpha_{#1}}}
\nc\efri[1]{F_{r \delta - \alpha_{#1}}}
\nc\efd[2][]{F_{#1 \delta,#2}}
\nc\efkd[1]{F_{k \delta,#1}}
\nc\efmd[1]{F_{m \delta,#1}}
\nc\efrd[1]{F_{r \delta,#1}}

\nc\fa{\frak a}
\nc\fb{\frak b}
\nc\fc{\frak c}
\nc\fd{\frak d}
\nc\fe{\frak e}
\nc\ff{\frak f}
\nc\fg{\frak g}
\nc\fh{\frak h}
\nc\fj{\frak j}
\nc\fk{\frak k}
\nc\fl{\frak l}
\nc\fm{\frak m}
\nc\fn{\frak n}
\nc\fo{\frak o}
\nc\fp{\frak p}
\nc\fq{\frak q}
\nc\fr{\frak r}
\nc\fs{\frak s}
\nc\ft{\frak t}
\nc\fu{\frak u}
\nc\fv{\frak v}
\nc\fz{\frak z}
\nc\fx{\frak x}
\nc\fy{\frak y}

\nc\fA{\frak A}
\nc\fB{\frak B}
\nc\fC{\frak C}
\nc\fD{\frak D}
\nc\fE{\frak E}
\nc\fF{\frak F}
\nc\fG{\frak G}
\nc\fH{\frak H}
\nc\fJ{\frak J}
\nc\fK{\frak K}
\nc\fL{\frak L}
\nc\fM{\frak M}
\nc\fN{\frak N}
\nc\fO{\frak O}
\nc\fP{\frak P}
\nc\fQ{\frak Q}
\nc\fR{\frak R}
\nc\fS{\frak S}
\nc\fT{\frak T}
\nc\fU{\frak U}
\nc\fV{\frak V}
\nc\fZ{\frak Z}
\nc\fX{\frak X}
\nc\fY{\frak Y}
\nc\tfi{\ti{\Phi}}
\nc\bF{\bold F}
\rnc\bol{\bold 1}

\nc\ua{\bold U_\A}

\def\cA{{\cal A}}   \def\cB{{\cal B}}   \def\cC{{\cal C}}
\def\cD{{\cal D}}   \def\cE{{\cal E}}   \def\cF{{\cal F}}  
\def\cG{{\cal G}}   \def\cH{{\cal H}}   \def\cI{{\cal I}}
\def\cJ{{\cal J}}   \def\cK{{\cal K}}   \def\cL{{\cal L}}
\def\cM{{\cal M}}   \def\cN{{\cal N}}   \def\cO{{\cal O}}
\def\cP{{\cal P}}   \def\cQ{{\cal Q}}   \def\cR{{\cal R}}
\def\cS{{\cal S}}   \def\cT{{\cal T}}   \def\cU{{\cal U}}
\def\cV{{\cal V}}   \def\cW{{\cal W}}   \def\cX{{\cal X}}
\def\cY{{\cal Y}}   \def\cZ{{\cal Z}}  \def\cRR{{\cal {\mathbb R}}}

\newcommand{\ben}{\begin{eqnarray}}
\newcommand{\een}{\end{eqnarray}}
\newcommand{\nonu}{\nonumber \\} 
\newcommand{\FF}{{\mathbb F}}
\newcommand{\A}{{\mathbb A}}
\newcommand{\GG}{{\mathbb G}}

\nc\qinti[1]{[#1]_i}
\nc\q[1]{[#1]_q}
\nc\xpm[2]{E_{#2 \delta \pm \alpha_#1}}  
\nc\xmp[2]{E_{#2 \delta \mp \alpha_#1}}
\nc\xp[2]{E_{#2 \delta + \alpha_{#1}}}
\nc\xm[2]{E_{#2 \delta - \alpha_{#1}}}
\nc\hik{\ed{k}{i}}
\nc\hjl{\ed{l}{j}}
\nc\qcoeff[3]{\left[ \begin{smallmatrix} {#1}& \\ {#2}& \end{smallmatrix}
\negthickspace \right]_{#3}}
\nc\qi{q}
\nc\qj{q}

\nc\ufdm{{_\ca\bu}_{\rm fd}^{\le 0}}


\nc\isom{\cong} 

\nc{\pone}{{\Bbb C}{\Bbb P}^1}
\nc{\pa}{\partial}
\def\H{\mathcal H}
\def\L{\mathcal L}
\nc{\F}{{\mathcal F}}
\nc{\Sym}{{\goth S}}
\nc{\arr}{\rightarrow}
\nc{\larr}{\longrightarrow}

\nc{\ri}{\rangle}
\nc{\lef}{\langle}
\nc{\W}{{\mathcal W}}
\nc{\uqatwoatone}{{U_{q,1}}(\su)}
\nc{\uqtwo}{U_q(\goth{sl}_2)}
\nc{\dij}{\delta_{ij}}
\nc{\divei}{E_{\alpha_i}^{(n)}}
\nc{\divfi}{F_{\alpha_i}^{(n)}}
\nc{\Lzero}{\Lambda_0}
\nc{\Lone}{\Lambda_1}
\nc{\ve}{\varepsilon}
\nc{\phioneminusi}{\Phi^{(1-i,i)}}
\nc{\phioneminusistar}{\Phi^{* (1-i,i)}}
\nc{\phii}{\Phi^{(i,1-i)}}
\nc{\Li}{\Lambda_i}
\nc{\Loneminusi}{\Lambda_{1-i}}
\nc{\vtimesz}{v_\ve \otimes z^m}

\nc{\asltwo}{\widehat{\goth{sl}_2}}
\nc\ag{\widehat{\goth{g}}}  
\nc\teb{\tilde E_\boc}
\nc\tebp{\tilde E_{\boc'}}

\title[]{Non-Abelian symmetries of the half-infinite XXZ spin chain}
\author{Pascal Baseilhac}
\address{Laboratoire de Math\'ematiques et Physique Th\'eorique CNRS/UMR 7350, F\'ed\'eration Denis Poisson FR2964,
F\'ed\'eration Denis Poisson, Universit\'e de Tours, Parc de Grammont, 37200 Tours, FRANCE}
\email{baseilha@lmpt.univ-tours.fr}

\author{Samuel Belliard}
\address{Institut de Physique Th\'eorique, Orme des Merisiers, batiment 774, Point courrier 136, CEA/DSM/IPhT, CEA/Saclay Gif-sur-Yvette Cedex FRANCE, 91191 Saint-Aubin}
\email{samuel.belliard@cea.fr}

\begin{abstract}The non-Abelian symmetries of the half-infinite XXZ spin chain for all possible types of integrable boundary conditions are classified. For each type of boundary conditions, an analog of the Chevalley-type presentation is given for the corresponding symmetry algebra. In particular, two new algebras arise that are, respectively, generated by the symmetry operators of the model with triangular and special $U_q(gl_2)-$invariant integrable boundary conditions.
\end{abstract}

\maketitle

\vskip -0.5cm

{\small MSC:\ 81R50;\ 81R10;\ 81U15.}

{{\small  {\it \bf Keywords}: Open XXZ spin chain, Coideal subalgebras, Non-Abelian symmetry, Quantum integrable models.}}
\vspace{0cm}

\section{Introduction}
Identifying non-Abelian infinite dimensional symmetries of quantum integrable systems provides a very efficient starting point for transferring the Hamiltonian and eigenstates formulation into the langage of operator algebra, representation theory and related special functions, on which the non-perturbative analysis is thus based. Besides the restricted class of models with conformal symmetry for which the Virasoro algebra and its representation theory play a central role, in the thermodynamic limit of lattice systems it is known that quantum groups, related current algebras and infinite dimensional $q-$vertex operators representations provide an appropriate mathematical framework. As an important example, in the thermodynamic limit $N\rightarrow \infty$ of the finite XXZ spin chain with $N$ sites and periodic boundary conditions, the $U_q(\widehat{sl_2})$ algebra emerges as a hidden non-Abelian symmetry of the Hamiltonian \cite{FM,Jim0,vertex}. Based on the representation theory of the $U_q(\widehat{sl_2})$ quantum affine algebra at level one and its current algebra, an explicit characterization of the Hamiltonian's spectrum, corresponding eigenstates, as well as multiple integral representations of correlation functions  and form factors of local operators has been given \cite{vertex}. This approach has been later on extended to lattice systems with periodic boundary conditions associated with higher spins  \cite{Idz} or higher rank affine Lie algebras \cite{Koy}, as well as for certain class of boundary conditions, see for instance \cite{JKKKM,BB3,BKo1,BKo2,Koj}. \vspace{1mm}

For lattice models with general integrable boundary conditions, identifying the non-Abelian symmetries in the thermodynamic limits has remained essentially unexplored although coideal subalgebras of quantum affine Lie algebras are natural candidates. For instance, in the thermodynamic limit  $N\rightarrow \infty$ of the {\it finite} open XXZ spin chain it was expected that non-Abelian infinite dimensional symmetries emerge, that are associated with certain coideal subalgebras of $U_q(\widehat{sl_2})$. Recall that the Hamiltonian of the half-infinite XXZ spin chain is formally defined as (see also \cite{JKKKM,BB3,BKo1,BKo2}):
\beqa
\qquad H_{\frac{1}{2}XXZ}&=&-\frac{1}{2}\sum_{k=1}^{\infty}\Big(\sigma_1^{k+1}\sigma_1^{k}+\sigma_2^{k+1}\sigma_2^{k} + \Delta\sigma_3^{k+1}\sigma_3^{k}\Big) + h_B  \label{Hsemi}
\eeqa
with  boundary interaction
\beqa
h_B= - \frac{(q-q^{-1})}{4}\frac{(\epsilon_+-\epsilon_-)}{(\epsilon_++\epsilon_-)}\sigma^1_3 - \frac{1}{(\epsilon_++\epsilon_-)}\big(k_+\sigma^1_+ + k_-\sigma^1_-\big).\nonumber
\eeqa
Here $\sigma_{1,2,3}$ and $\sigma_\pm=(\sigma_1\pm i\sigma_2)/2$ are usual Pauli matrices\footnote{\beqa
\sigma_+=\left(
\begin{array}{cc}
 0    & 1 \\
 0 & 0 
\end{array} \right) \ ,\qquad
\sigma_- =\left(
\begin{array}{cc}
 0    & 0 \\
 1 & 0 
\end{array} \right) \ ,\qquad
\sigma_3 =\left(
\begin{array}{cc}
 1    & 0 \\
 0 & -1 
\end{array} \right) \ .\label{Pauli}\nonumber
\eeqa
}, $\Delta=(q+q^{-1})/2$\ \ denotes the anisotropy parameter and $\epsilon_\pm,k_\pm$  are scalar parameters associated with the right boundary field. Formally, the Hamiltonian  acts on an infinite tensor product of $2-$dimensional  vector spaces. Note that the ordering of the tensor components in (\ref{Hsemi}) is such that:
\beqa
{\cal V}= \cdot \cdot\cdot \otimes {\mathbb C}^2 \otimes {\mathbb C}^2  \otimes {\mathbb C}^2 \ .\label{halfV}
\eeqa
Given the Hamiltonian of the half-infinite XXZ spin chain (\ref{Hsemi}), until recently the hidden non-Abelian symmetry  for $q$ generic and any type of boundary conditions\footnote{In (\ref{Hsemi}), we implicitly assume that $\epsilon_++\epsilon_-\neq 0$.} $k_\pm,\epsilon_\pm$ has remained unknown. However, at the end of the article \cite{BB3}, it was pointed out that for the generic case ($k_\pm\neq 0$ and $\epsilon_\pm\neq 0$) and the diagonal case ($k_\pm=0$ and $\epsilon_\pm\neq 0$) the respective hidden non-Abelian symmetries are associated with two different coideal subalgebras of $U_q(\widehat{sl_2})$: the $q$-Onsager algebra \cite{Ter03,Bas1} and the augmented $q$-Onsager algebra \cite{BC12,BB3} (see also \cite{IT}). 
\vspace{1mm}

In this letter, we present a unified picture that complete the preliminary observations of \cite{BB3}. Namely, for each type of boundary conditions, the non-Abelian symmetry algebra of the half-infinite XXZ spin chain 
is identified and characterized through generators and relations. In particular, two new coideal subalgebras are obtained.  
\vspace{1mm} 

The text is organized as follows. In Section 2, four different types of coideal subalgebras of $U_q(\widehat{sl_2})$ are defined though generators and relations in a Chevalley-type presentation. The coaction and counit maps are given in each case. Then, it is shown that these algebras are symmetry algebras of the Hamiltonian (\ref{Hsemi}) for generic ($\epsilon_\pm\neq 0,k_\pm\neq 0$), triangular ($\epsilon_\pm\neq 0,k_+= 0,k_-\neq 0$), diagonal ($\epsilon_\pm\neq 0,k_\pm= 0$) and special\footnote{The corresponding Hamiltonian can be understood as the thermodynamic limit of the $U_q(sl_2)$ invariant spin chain studied in \cite{PS}.}
 boundary conditions ($\epsilon_+=1, \epsilon_-=0,k_\pm\neq 0$), respectively. Note that since the case of upper triangular ($k_-=0$, $k_+\neq 0$) and lower triangular ($k_+=0$, $k_-\neq 0$) boundary conditions  are related through
conjugation of the Hamiltonian (\ref{Hsemi}) by the spin-reversal operator $\hat{\nu}=\prod_{j=1}^\infty \sigma_j^x$, it is sufficient for our purpose to restrict our attention to the case of lower boundary conditions. In Section 3, we propose an alternative and simpler derivation of the symmetry operators which is based on the remarkable connection between the infinite dimensional algebra ${\cal A}_q$ introduced in \cite{BK1} and the $q-$Onsager algebra. Concluding remarks follow in Section 4.
\vspace{1mm}    

\begin{notation}
The $q-$commutator $\big[X,Y\big]_q=qXY-q^{-1}YX$ is introduced, where $q$ is the deformation parameter assumed not to be a root of unity.
\end{notation}

\section{Four different types of $q-$Onsager symmetry algebras}
In the first part of this Section, four different types of coideal subalgebras of $U_{q}(\widehat{sl_2})$ with central extension are introduced through basic generators and relations. The coaction and counit maps\footnote{In general, given a Hopf algebra ${\cal H}$ with comultiplication $\Delta$ and counit ${\cal E}$, ${\cal I}$ is
called a left ${\cal H}-$comodule (coideal subalgebra of ${\cal H}$) if there exists a 
coaction map $\delta:\ \ {\cal I}\rightarrow {\cal H}
\otimes {\cal I}$ such that (right coaction maps are defined similarly)
\beqa (\Delta \times id)\circ
\delta=(id\times\delta)\circ\delta\ ,\qquad
({\cal E} \times id)\circ \delta \cong id \
.\label{defcoaction}\nonumber\eeqa
 } are given. Note that two of these algebras have already appeared in the literature: the $q-$Onsager algebra \cite{Ter03,Bas1} and the augmented $q-$Onsager  algebra \cite{IT,BC12,BB3}. The two others, the so-called triangular $q-$Onsager algebra and $U_q(gl_2)$ invariant $q-$Onsager algebra\footnote{However, note that the generators of the $U_q(gl_2)$ invariant $q$-Onsager algebra have already appeared in \cite{Vidas,Kolb}.} are new. In a second part, depending on the choice of boundary conditions, by analogy with \cite[eq. (3.2)]{Jim0} it is shown that the Hamiltonian (\ref{Hsemi}) commutes with all generators of one of these four algebras.
 
 Define the extended Cartan matrix $\{a_{ij}\}$ ($a_{ii}=2$,\ $a_{ij}=-2$ for $i\neq j$). The quantum affine algebra $U_{q}(\widehat{sl_2})$ is generated by the elements $\{h_j,e_j,f_j\}$, $j\in \{0,1\}$ which satisfy the defining relations
 \beqa [h_i,h_j]=0\ , \quad [h_i,e_j]=a_{ij}e_j\ , \quad
 [h_i,f_j]=-a_{ij}f_j\ ,\quad
 [e_i,f_j]=\delta_{ij}\frac{q^{h_i}-q^{-h_i}}{q-q^{-1}}\
 \nonumber\eeqa
 together with the $q-$Serre relations
 \beqa [e_i,[e_i,[e_i,e_j]_{q}]_{q^{-1}}]=0\ ,\quad \mbox{and}\quad
 [f_i,[f_i,[f_i,f_j]_{q}]_{q^{-1}}]=0\ . \label{defUq}\eeqa
 The sum $ c=h_0+h_1$ is the central element of the algebra. The
 Hopf algebra structure is ensured by the existence of a
 comultiplication $\Delta: U_{q}(\widehat{sl_2})\mapsto U_{q}(\widehat{sl_2})\otimes U_{q}(\widehat{sl_2})$, antipode ${\cal S}: U_{q}(\widehat{sl_2})\mapsto U_{q}(\widehat{sl_2})$ 
 and a counit ${\cal E}: U_{q}(\widehat{sl_2})\mapsto {\mathbb C}$ with
 \beqa \Delta(e_i)&=&e_i\otimes 1 +
 q^{h_i}\otimes e_i\ ,\nonumber \\
  \Delta(f_i)&=&f_i\otimes q^{-h_i} + 1 \otimes f_i\ ,\nonumber\\
  \Delta(h_i)&=&h_i\otimes 1 + 1 \otimes h_i\ ,\label{coprod}
 \eeqa
 \beqa {\cal S}(e_i)=-q^{-h_i}e_i\ ,\quad {\cal S}(f_i)=-f_iq^{h_i}\ ,\quad {\cal S}(h_i)=-h_i \qquad {\cal S}({1})=1\
 \label{antipode}\nonumber\eeqa
 and\vspace{-0.3cm}
 \beqa {\cal E}(e_i)={\cal E}(f_i)={\cal
 E}(h_i)=0\ ,\qquad {\cal E}(1)=1\
 .\label{counit}\nonumber\eeqa
 Note that the opposite coproduct $\Delta'$ can be similarly defined with $\Delta'\equiv \sigma
 \circ\Delta$ where the permutation map $\sigma(x\otimes y
 )=y\otimes x$ for all $x,y\in U_{q}(\widehat{sl_2})$ is used.\vspace{2mm} 

\subsection{The $q-$Onsager algebra}
The $q-$Onsager algebra $O_q(\widehat{sl_2})$ is an example of tridiagonal algebra \cite{Ter03}. Including a central extension, it is generated by two elements ${\textsf W}_0,{\textsf W}_1$, a central element $\Gamma$ and unit. The defining relations are:
\beqa \big[{\textsf W}_0,\big[{\textsf W}_0,\big[{\textsf W}_0,{\textsf
W}_1\big]_q\big]_{q^{-1}}\big]&=&\rho\big[{\textsf W}_0,{\textsf W}_1\big]\ ,\label{Talggen}\\
\big[{\textsf W}_1,\big[{\textsf W}_1,\big[{\textsf W}_1,{\textsf
W}_0\big]_q\big]_{q^{-1}}\big]&=&\rho\big[{\textsf W}_1,{\textsf W}_0\big], \nonumber\\
\big[{\textsf W}_0,\Gamma\big]=\big[{\textsf W}_1,\Gamma\big]=0,\nonumber
\eeqa
where $\rho$ is a scalar. Without loss of generality, let us define
\beqa
\rho =(q+q^{-1})^2 k_+k_-\ \label{rho}
\eeqa
where $k_\pm$ are nonzero scalars. By analogy with the situation for Hopf algebras, one endows the $q-$Onsager algebra with the coaction map   $\delta: O_q(\widehat{sl_2}) \rightarrow U_q(\widehat{sl_2}) \otimes O_q(\widehat{sl_2})$  defined by:
\beqa
\delta({\textsf W}_0)&=& (k_+e_1 + k_-q^{-1}f_1q^{h_1})\otimes 1 + q^{h_1} \otimes {\textsf W}_0\ ,\label{deltadef}\\
\delta({\textsf W}_1)&=& (k_-e_0 + k_+q^{-1}f_0q^{h_0})\otimes 1 + q^{h_0} \otimes {\textsf W}_1\ ,\nonumber\\
\delta(\Gamma)&=&q^c\otimes \Gamma\nonumber
\eeqa
and counit ${\cal E}:  O_q(\widehat{sl_2}) \rightarrow {\mathbb C}$: 
\beqa
{\cal E}({\textsf W}_0)=\epsilon_+, \quad {\cal E}({\textsf W}_1)=\epsilon_-, \quad {\cal E}(\Gamma)=  {\cal E}(1)=1. 
\eeqa

This induces an homomorphism $\psi=(id \times {\cal E})\circ \delta: O_q(\widehat{sl_2})\rightarrow U_q(\widehat{sl_2}) $ from the $q-$Onsager algebra to a subalgebra of $U_q(\widehat{sl_2})$:
\beqa
\psi({\textsf W}_0)&=& k_+e_1 + k_-q^{-1}f_1q^{h_1} + \epsilon_+ q^{h_1} \ ,\label{realop}\\
\psi({\textsf W}_1)&=& k_-e_0 + k_+q^{-1}f_0q^{h_0} + \epsilon_-q^{h_0}, \nonumber\\
\psi(\Gamma)&=&q^c.\nonumber
\eeqa
\vspace{1mm}

\subsection{The triangular $q-$Onsager algebra}
The triangular $q-$Onsager algebra $O^{t}_q(\widehat{sl_2})$ is generated by three elements ${\textsf T}_0,{\textsf T}_1$, $\tilde{\textsf{P}}_1$, a central element $\Gamma$ and unit. The defining relations are:
\beqa
\big[{\textsf T}_0,\big[{\textsf T}_0,\tilde{\textsf P}_1\big]_{q^{-1}}\big]&=&\rho_{t}\big[{\textsf T}_0,{\textsf T}_1\big],\qquad \qquad \qquad
\big[{\textsf T}_1,\big[{\textsf T}_1,\tilde{\textsf P}_1\big]_{q}\big]=\rho_{t}\big[{\textsf T}_0,{\textsf T}_1\big],\label{Ttrianggen}\\
\big[{\textsf T}_1,{\textsf
T}_0\big]_{q^{-1}}&=& \tilde\rho_t\Gamma,\qquad \qquad \qquad \qquad \big[{\textsf T}_0,\Gamma\big]=\big[{\textsf T}_1,\Gamma\big]= \big[\tilde{\textsf P}_1,\Gamma\big]=0.\nonumber
\eeqa
Let us define:
\beqa
\rho_{t}=k_-(q+q^{-1})^2 \qquad \mbox{and} \qquad \tilde\rho_{t}=-\epsilon_+\epsilon_-(q-q^{-1}),\label{rhot}
\eeqa
where $k_-,\epsilon_\pm$ are nonzero scalars. We endow this algebra with the coaction map   $\delta_t:  O^{t}_q(\widehat{sl_2}) \rightarrow U_q(\widehat{sl_2}) \otimes O^{t}_q(\widehat{sl_2})$  defined by:
\beqa
\delta_t({\textsf T}_0)&=& k_-q^{-1}f_1q^{h_1} \otimes 1 + q^{h_1} \otimes {\textsf T}_0\ ,\label{deltadeft}\\
\delta_t({\textsf T}_1)&=& k_-e_0 \otimes 1 + q^{h_0} \otimes {\textsf T}_1\ ,\nonumber\\
\delta_t(\tilde{\textsf P}_1)&=&  k_-\big([e_1,e_0]_q+q^c [f_1,f_0]_q\big) \otimes 1 + q^c \otimes \tilde{\textsf P}_1 + 
(q^2-q^{-2})\left( q^{-1} e_1q^{h_0} \otimes {\textsf T}_1  +  q^c f_0 q^c \otimes {\textsf T}_0 \right),\nonumber\\
\delta_t(\Gamma)&=&q^c\otimes \Gamma\nonumber
\eeqa
and counit ${\cal E}_t:  O^t_q(\widehat{sl_2}) \rightarrow {\mathbb C}$: 
\beqa
{\cal E}_t({\textsf T}_0)=\epsilon_+, \quad {\cal E}_t({\textsf T}_1)=\epsilon_-, \quad {\cal E}_t(\tilde{\textsf P}_1)=\tilde p, \quad {\cal E}_t(\Gamma)={\cal E}_t(1)=1. 
\eeqa
This induces an homomorphism $\psi_t=(id \times {\cal E}_t)\circ \delta_t$ from the triangular $q-$Onsager algebra to a subalgebra of $U_q(\widehat{sl_2})$:
\beqa
\psi_t({\textsf T}_0)&=& k_-q^{-1}f_1q^{h_1} + \epsilon_+ q^{h_1},\label{realopt}\\
\psi_t({\textsf T}_1)&=&k_-e_0 + \epsilon_-q^{h_0},\nonumber\\
\psi_t(\tilde{\textsf P}_1)&=& (q^2-q^{-2})\big( \epsilon_-q^{-1} e_1q^{h_0}  + \epsilon_+ f_0 q^{h_1+h_0}\big)+k_-\big([e_1,e_0]_q+q^c [f_1,f_0]_q\big)+\tilde p q^c, \nonumber\\
\psi_t(\Gamma)&=&q^c.\nonumber
\eeqa
\vspace{1mm}

\subsection{The augmented $q-$Onsager algebra}
The augmented $q-$Onsager algebra $O^{d}_q(\widehat{sl_2})$ has been introduced in \cite{BB3}, as a generalization\footnote{In \cite{IT}, the special case $\Gamma=1$ is considered. In \cite{BB3}, the central element is not explicitly introduced and the first relation is replaced by $[{\textsf K}_0, {\textsf K}_1]=0$. Note that for the level one $q-$vertex operators representations constructed in \cite{BB3}, one has $\Gamma=q$.} of the augmented tridiagonal algebra introduced in \cite{IT}. It is generated by four elements ${\textsf K}_0,{\textsf K}_1, {\textsf Z}_1, \tilde{\textsf Z}_1$, a central element $\Gamma$ and unit. The defining relations are:
\beqa
{\textsf K}_0 {\textsf K}_1&=&{\textsf K}_1 {\textsf K}_0=\epsilon_-\epsilon_+\Gamma\ ,\label{Tauggen}\\
 {\textsf K}_0{\textsf Z}_1&=&q^{-2} {\textsf Z}_1{\textsf K}_0\ ,\qquad {\textsf K}_0\tilde{\textsf Z}_1=q^{2}\tilde{\textsf Z}_1{\textsf K}_0\ ,\nonumber\\
{\textsf K}_1{\textsf Z}_1&=&q^{2}{\textsf Z}_1{\textsf K}_1\ ,\ \ \qquad{\textsf K}_1\tilde{\textsf Z}_1=q^{-2}\tilde{\textsf Z}_1{\textsf K}_1\ ,\nonumber\\
\big[{\textsf Z}_1,\big[{\textsf Z}_1,\big[{\textsf Z}_1,\tilde{\textsf
Z}_1\big]_q\big]_{q^{-1}}\big]&=&\rho_{d}{\textsf Z}_1(\,{\textsf K}_1{\textsf K}_1-\,{\textsf K}_0{\textsf K}_0){\textsf Z}_1,\nonumber\\
\big[\tilde{\textsf Z}_1,\big[\tilde{\textsf Z}_1,\big[\tilde{\textsf Z}_1,{\textsf
Z}_1\big]_q\big]_{q^{-1}}\big]&=&\rho_{d}\tilde{\textsf Z}_1({\textsf K}_0{\textsf K}_0-{\textsf K}_1{\textsf K}_1)\tilde{\textsf Z}_1,\  \nonumber\\
&& \!\!\!\!\!\!\!\!\!\!\!\!\!\!\!\!\!\!\!\!\!\!\!\!\!
\!\!\!\!\!\!\!\!\!\!\!\!\!\!\!\!\!\!\!\!\!\!\!\!\!\!\!\!\!\!\!\!\!\!\!\!\!\!\!\!\!\! \big[{\textsf Z}_1,\Gamma\big]=\big[\tilde{\textsf Z}_1,\Gamma\big]=\big[{\textsf K}_0,\Gamma\big]= \big[{\textsf K}_1,\Gamma\big]=0\nonumber
\eeqa
with the identification:
\ben
\rho_{d}=\frac{(q^3-q^{-3})(q^2-q^{-2})^3}{q-q^{-1}}.\label{rhod}
\een

We endow this algebra with the coaction map   $\delta_d: O^{d}_q(\widehat{sl_2}) \rightarrow U_q(\widehat{sl_2}) \otimes O^{d}_q(\widehat{sl_2})$  defined by:
\beqa
\delta_d({\textsf K}_0)&=& q^{h_1} \otimes {\textsf K}_0\ ,\qquad \delta_d({\textsf K}_1)= q^{h_0} \otimes {\textsf K}_1\ ,\label{deltadefaug}\\
\delta_d({\textsf Z}_1)&=& q^{c} \otimes {\textsf Z}_1 + (q^2-q^{-2})\big( q^{-1} e_0q^{h_1} \otimes {\textsf K}_0 +  f_1q^{c} \otimes {\textsf K}_1\big) \ ,\nonumber\\
\delta_d(\tilde{\textsf Z}_1)&=&   q^{c} \otimes \tilde{\textsf Z}_1 + (q^2-q^{-2})\big( f_0q^{c} \otimes {\textsf K}_0 + q^{-1} e_1q^{h_0} \otimes {\textsf K}_1\big) \ ,\nonumber\\
\delta_d(\Gamma)&=&q^c\otimes \Gamma\nonumber
\eeqa
and counit ${\cal E}_d:  O^d_q(\widehat{sl_2}) \rightarrow {\mathbb C}$: 
\beqa
{\cal E}_d({\textsf K}_0)=\epsilon_+, \quad {\cal E}_d({\textsf K}_1)=\epsilon_-, \quad     {\cal E}_d({\textsf Z}_1)={\cal E}_d(\tilde{\textsf Z}_1)=0,\quad {\cal E}_d(\Gamma)=\epsilon_-\epsilon_+,\quad  {\cal E}_d(1)=1. 
\eeqa
This induces an homomorphism $\psi_d$ from the augmented $q-$Onsager algebra to a subalgebra of $U_q(\widehat{sl_2})$:
\beqa
\psi_d({\textsf K}_0)&=& \epsilon_+q^{h_1}  \ ,\qquad \psi_d({\textsf K}_1)= \epsilon_- q^{h_0} \ ,\label{realopd} \\
\psi_d({\textsf Z}_1)&=&      (q^2-q^{-2})\big( \epsilon_+q^{-1} e_0q^{h_1}  + \epsilon_- f_1 q^{h_1+h_0}\big) \ ,\nonumber\\
\psi_d(\tilde{\textsf Z}_1)&=& (q^2-q^{-2})\big( \epsilon_-q^{-1} e_1q^{h_0}  + \epsilon_+ f_0 q^{h_1+h_0}\big), \nonumber\\
\psi_d(\Gamma)&=&\epsilon_-\epsilon_+q^c.\nonumber
\eeqa

\subsection{The $U_q(gl_2)$ invariant $q-$Onsager algebra:} The  $U_q(gl_2)$ invariant $q-$Onsager algebra $O^{i}_q(\widehat{sl_2})$ is generated by six elements ${\textsf e},{\textsf f}, q^{\textsf h}, {\textsf X}, {\textsf Y}, \tilde{\textsf Y}$, a central element $\Gamma$ and unit. The defining relations are:
\ben
\quad\quad[{\textsf e},{\textsf f}]=\frac{q^{\textsf h}-q^{-{\textsf h}}}{q-q^{-1}}, \quad [{\textsf e},q^{\textsf h}]_q=[q^{\textsf h},{\textsf f}]_q=0,
\een
\ben
&&[q^{\textsf h},{\textsf X}]=0, \quad [{\textsf X},{\textsf e}]={\textsf Y},\quad [{\textsf f},{\textsf X}]= \tilde{\textsf Y},\nonumber
\een
\ben
&&[q^{\textsf h},\tilde{\textsf Y}]_q=0, \quad [{\textsf e},{\textsf Y}]_q=0, \quad [q^{\textsf h},{\textsf Y}]_{q^{-1}}=0,\quad [{\textsf f},\tilde{\textsf Y}]_{q^{-1}}=0,\nonumber
\een
\ben
[{\textsf e},\tilde{\textsf Y}]=(q+q^{-1}){\textsf X}q^{\textsf h},\quad
[{\textsf Y},{\textsf f}]=(q+q^{-1}){\textsf X}q^{\textsf h},\nonumber
\een
\ben
 [{\textsf X},[{\textsf Y},\tilde{\textsf Y}]]&=&(q^2-q^{-2})^3(q\tilde{\textsf Y}{\textsf f}{\textsf e}{\textsf e}-q^{-1}{\textsf Y}{\textsf e}{\textsf f}{\textsf f}+(q+q^{-1})^2 {\textsf X}{\textsf e}{\textsf f}q^{\textsf h})q^{\textsf h}\nonumber\\ &&+\ q^2(q^2-q^{-2})(q+q^{-1})^2\Big((q^2-q^{-2}-1)\tilde{\textsf Y}{\textsf e}+(q^2-q^{-2}+1){\textsf Y}{\textsf f}-(q+q^{-1}) {\textsf X}q^{\textsf h} \Big)q^{2\textsf h}\nonumber\\
 &&+\ (q^2-q^{-2})(q+q^{-1})^2\Big(q^{-2}\tilde{\textsf Y}{\textsf e}-q^{2}{\textsf Y}{\textsf f}+q^2(q+q^{-1}){\textsf X}q^{\textsf h}\Big),\nonumber\\
&& \big[{\textsf Y},\Gamma\big]=\big[\tilde{\textsf Y},\Gamma\big]=\big[{\textsf X},\Gamma\big]= \big[{\textsf e},\Gamma\big]= \big[{\textsf f},\Gamma\big]= \big[{\textsf q^{h}},\Gamma\big]=0.\nonumber
\een

We endow this algebra with the coaction map   $\delta_i: O^{i}_q(\widehat{sl_2}) \rightarrow U_q(\widehat{sl_2}) \otimes O^{i}_q(\widehat{sl_2})$  defined by:
\beqa
\delta_i({\textsf e})&=& e_0 \otimes 1 +
q^{h_0}\otimes {\textsf e}, \qquad \delta_i({\textsf f})= f_0 \otimes q^{-{\textsf h}} +
1 \otimes {\textsf f}, \qquad  \delta_i(q^{\textsf h}) = q^{h_0} \otimes q^{\textsf h},\label{deltadefi}\\
\delta_i({\textsf X})&=&([e_1,e_0]_q-[f_1,f_0]_{q^{-1}} q^{c})\otimes 1+q^{c}\otimes {\textsf X}+(q^{2}-q^{-2})(q^{h_0+1} e_1 \otimes  {\textsf e}- q^{c-1} f_1 \otimes {\textsf f} q^{\textsf h}),\nonumber\\
\delta_i({\textsf Y})&=&(\big[\big[e_1,e_0\big],e_0\big]_{q}-q^{c+1}(q+q^{-1})f_1q^{h_0})\otimes 1 + q^{h_0+c}\otimes {\textsf Y}\\
&&+(q^2-q^{-2})\Big(q^{h_0}[e_1,e_0]_q \otimes {\textsf e}+(q^2-1) q^{2 h_0} e_1\otimes {\textsf e}^2+ q^{c+h_0} f_1 \otimes \frac{q^{2  {\textsf h}}-1}{q^{2}-1} \Big) ,\nonumber\\
\delta_i(\tilde{\textsf Y})&=& q^{c}\big(\big[\big[f_1,f_0\big],f_0\big]_{q^{-1}}-q^{-1}(q+q^{-1})e_1q^{h_0}\big)\otimes q^{-{\textsf h}}+q^{c}\otimes \tilde{\textsf Y}-(q^2-q^{-2})\Big( q^{c}[f_0,f_1]_q \otimes  {\textsf f} \\
&&+q^{1+h_0} e_1 \otimes  \frac{q^{\textsf h}-q^{-\textsf h}}{q-q^{-1}}+(q-q^{-1}) q^c f_1 \otimes {\textsf f}^2 q^{-{\textsf h}}\Big ),\nonumber\\
\delta_i(\Gamma)&=&q^c\otimes \Gamma\nonumber
\eeqa
and counit ${\cal E}_i:  O^i_q(\widehat{sl_2}) \rightarrow {\mathbb C}$:
\beqa
{\cal E}_i(e)={\cal E}_i(f)={\cal E}_i({\textsf X})={\cal E}_i({\textsf Y})={\cal E}_i(\tilde{\textsf Y})= 0, \quad {\cal E}_i(q^{h})={\cal E}_i(\Gamma)={\cal E}_i(1)=1. 
\eeqa

This induces an homomorphism  $\psi_i=(id \otimes  {\cal E}_i)\circ \delta_i $ from the $U_q(gl_2)$ invariant $q-$Onsager algebra to a certain subalgebra of $U_q(\widehat{sl_2})$:
\ben
&&\psi_i({\textsf e})= e_0, \quad \psi_i({\textsf f}) = f_0 , \quad \psi_i(q^{\textsf h})= q^{h_0},\label{realopi}\\
&&\psi_i({\textsf X})= [e_1,e_0]_q-[f_1,f_0]_{q^{-1}} q^{c},\nonumber\\
&&\psi_i({\textsf Y})= \big[\big[e_1,e_0\big],e_0\big]_{q}-q^{c+1}(q+q^{-1})f_1q^{h_0},\nonumber\\
&&\psi_i(\tilde {\textsf Y}) = q^{c}\big(\big[\big[f_1,f_0\big],f_0\big]_{q^{-1}}-q^{-1}(q+q^{-1})e_1q^{h_0}\big),\nonumber\\
&&\psi_i(\Gamma)=q^c.\nonumber
\een
\vspace{1mm}


\subsection{Non-Abelian symmetry algebras of the Hamiltonians}
In the thermodynamic limit of the XXZ spin chain Hamiltonian, it is well-known that the Hamiltonian commutes with the generators of the quantum affine algebra $U_q(\widehat{sl_2})$ acting on an infinite tensor product representation \cite{FM,Jim0,vertex}.  In this subsection, we basically use similar arguments in order to characterize the hidden non-Abelian symmetries of the open XXZ spin chain for four different types of boundary conditions. As an example, let us consider the Hamiltonian (\ref{Hsemi}). On the semi-infinite tensor product vector space (\ref{halfV}), the generators (\ref{realop}) of the $q-$Onsager algebra act as\footnote{The (evaluation in the principal gradation) endomorphism  $\pi_\zeta: U_{q}(\widehat{sl_2}) \mapsto \mathrm{End}({\cal V}_\zeta)$ $({\cal V}\equiv{\mathbb C}^2)$ is used:
\beqa 
&&\pi_\zeta[e_1]= \zeta\sigma_+\ , \qquad \ \ \ \ \ \pi_\zeta[e_0]= \zeta\sigma_-\ , \nonumber\\
&&\pi_\zeta[f_1]=
\zeta^{-1}\sigma_-\ ,\qquad \pi_\zeta[f_0]= \zeta^{-1}\sigma_+\ ,\nonumber\\
\ &&\pi_\zeta[q^{h_1}]= q^{\sigma_3}\ ,\quad \qquad  \ \pi_\zeta[q^{h_0}]=
q^{-\sigma_3}\ .\ \label{evalrep}
\eeqa
To obtain (\ref{opfund}), one fixes the evaluation  parameter $\zeta=1$.
} (see also \cite{BB3}):
 \beqa
 \qquad \qquad {\cal W}^{(\infty)}_0&=& \sum_{j=1}^{\infty}\big(   \cdots\otimes q^{\sigma_3} \otimes   q^{\sigma_3}\otimes \underbrace{(k_+\sigma_+ + k_-\sigma_-)}_{site j} \otimes  I\!\!I \otimes  \cdots \otimes I\!\!I  \big)\quad + \quad \epsilon_+  \  \big( \cdots\otimes q^{\sigma_3} \otimes q^{\sigma_3} \big), \label{opfund} \\
 \qquad\qquad  {\cal W}^{(\infty)}_1&=& \sum_{j=1}^{\infty}  \big(  \cdots\otimes q^{-\sigma_3} \otimes   q^{-\sigma_3}\otimes \underbrace{(k_+\sigma_+ + k_-\sigma_-)}_{site j} \otimes  I\!\!I \otimes  \cdots \otimes  I\!\!I  \big)\quad + \quad  \epsilon_-  \ \big(   \cdots\otimes q^{-\sigma_3} \otimes q^{-\sigma_3}\big), \nonumber
 \eeqa
where the coaction map (\ref{deltadef}) is iterated repeatedly and (\ref{realop}) is used.  The action of the operators (\ref{opfund}) on (\ref{Hsemi}) is easy to compute. On one hand, introduce the local density $h_i=\sigma_1^{i+1}\sigma_1^{i}+\sigma_2^{i+1}\sigma_2^{i} + \Delta\sigma_3^{i+1}\sigma_3^{i}$. By straightforward calculations, one first observes:
\beqa
\big[h_i,{\cal W}^{(\infty)}_0\big] =   \cdots \otimes q^{\sigma_3} \otimes   q^{\sigma_3}\otimes \Gamma_{i+1,i} \otimes  I\!\!I \otimes  \cdots \otimes I\!\!I \ \label{commuthi}
\eeqa
where 
\beqa
\Gamma_{i+1,i}  = (2k_+\sigma_+  - 2k_-\sigma_-)   \otimes (\Delta -q^{\sigma_3} )\sigma_3 +   (\Delta q^{\sigma_3} - 1)\sigma_3   \otimes (2k_+\sigma_+  - 2k_-\sigma_-) .\nonumber 
\eeqa
Summing up the local densities, after some simplifications one ends up with:
\beqa
\big[ \sum_{i=1}^{\infty}h_i,{\cal W}^{(\infty)}_0\big] = -\frac{1}{2}(q-q^{-1})      \big( \cdots\otimes q^{\sigma_3} \otimes q^{\sigma_3} \big) \otimes (k_+\sigma_+  - k_-\sigma_-) .\nonumber
\eeqa

On the other hand, the contribution from the boundary term in (\ref{Hsemi}) is non-vanishing. It gives:
\beqa
\big[h_B,{\cal W}^{(\infty)}_0\big]= \frac{1}{2}(q-q^{-1})   \big( \cdots\otimes q^{\sigma_3} \otimes q^{\sigma_3} \big) \otimes (k_+\sigma_+  - k_-\sigma_-).\nonumber
\eeqa

Combining the above expressions together and repeating the analysis for ${\cal W}^{(\infty)}_1$,   we conclude that the $q-$Onsager algebra is a symmetry algebra of the Hamiltonian with generic boundary conditions:
\beqa
\big[ H_{\frac{1}{2}XXZ},{\cal W}^{(\infty)}_0\big]=0\quad \mbox{and}\quad \big[ H_{\frac{1}{2}XXZ},{\cal W}^{(\infty)}_1\big]=0.\label{Hcom}
\eeqa
\vspace{1mm}

Although the calculations become quickly more involved, the same technique is then extended to the Hamiltonian  (\ref{Hsemi})  with triangular ($\epsilon_\pm\neq 0,k_+= 0,k_-\neq 0$), diagonal ($\epsilon_\pm\neq 0,k_\pm= 0$) and special boundary conditions ($\epsilon_+=1, \epsilon_-=0,k_\pm\neq 0$), respectively. With respect to the boundary conditions chosen, it is found that the corresponding Hamiltonian  is commuting with all generators of the triangular $q-$Onsager, augmented $q-$Onsager and $U_q(gl_2)$ invariant $q-$Onsager algebras acting on (\ref{halfV}), respectively. In the next Section, we present an alternative and much simpler derivation of the symmetry operators generating the four different types of $q-$Onsager algebras.  \vspace{1mm}

\section{An alternative derivation of the symmetry operators}
In the $q-$Onsager approach of the  {\it finite} open XXZ spin chain  \cite{BK2} with left and right {\it generic non-diagonal} boundary conditions, it is known that all {\it local} or {\it non-local} conserved quantities are generated from elements in a certain quotient\footnote{Because the generators of ${\cal A}_q$ act on a finite dimensional  vector space ${\cal V}^{(N)}$ of dimension $2^N$,
they satisfy additionnal relations that are $q-$deformed analogs of Davies' relations. Compare the first reference of \cite[eqs. (17),(18)]{BK2} to
the first reference of \cite[eq. (2.6a),(2.6b)]{Davies} for details.}  of the infinite dimensional algebra ${\cal A}_q$ \cite{BK2} (see also \cite{BK1}). In the literature, the generators of ${\cal A}_q$ are usually denoted $\{{\cW}_{-k},{\cW}_{k+1},$ ${\cG}_{k+1},{\tilde{\cG}}_{k+1}|k\in{\mathbb Z}_+\}$, and ${\cW}_{0},{\cW}_{1}$ are called the fundamental generators. Importantly, ${\cW}_{0},{\cW}_{1}$ satisfy the defining relations of the $q-$Onsager algebra (\ref{Talggen}) \cite{Bas1,Bas2}.\vspace{1mm}

By analogy with Onsager's \cite{Ons44}, Dolan-Grady's \cite{DG} and Davies' \cite{Davies} works on the Onsager algebra, generalizing the results of \cite[Subsection 2.4.2]{BK1} it was shown in \cite[Example 2]{BB1} through a brute force calculation that the first few `descendants' generators  ${\cal W}_{-1},{\cal W}_{2},{\cal G}_{1},{\tilde{\cal G}}_{1},{\cal G}_{2},{\tilde{\cal G}}_{2}$ can be written as polynomials in the two fundamental generators  ${\cal W}_0,{\cal W}_1$ only. Including a central element $\gamma$, for instance one finds:
\beqa
&&{\cal{G}}_{1} = \big[{\cal{W}_1},\cal{W}_0\big]_q-(q-q^{-1})\epsilon_+\epsilon_-\gamma
 \ ,\label{defel}\\
&&{\cal{W}}_{-1} = \frac{1}{\rho}\left( (q^2+q^{-2})\cal{W}_0\cal{W}_1\cal{W}_0 -\cal{W}_0^2\cal{W}_1 - \cal{W}_1 \cal{W}_0^2\right) + \cal{W}_1-\frac{(q-q^{-1})^2}{\rho}\epsilon_+\epsilon_-\gamma  \cal{W}_0
\ ,\nonumber\\
&&{\cal{G}}_{2} = \frac{1}{\rho(q^2+q^{-2})} \Big( (q^{-3}+q^{-1}) \cal{W}_0^2{\cal{W}_1}^2 - (q^{3}+q){\cal{W}_1}^2\cal{W}_0^2 + (q^{-3}-q^{3})(\cal{W}_0{\cal{W}_1}^2\cal{W}_0 + {\cal{W}_1}\cal{W}_0^2{\cal{W}_1})  \nonumber\\
&&\qquad \qquad  - (q^{-5}+q^{-3} +2q^{-1}) \cal{W}_0{\cal{W}_1}\cal{W}_0{\cal{W}_1} + (q^{5}+q^{3} +2q){\cal{W}_1}\cal{W}_0{\cal{W}_1}\cal{W}_0  \nonumber\\
&&\qquad \qquad \qquad +\rho(q-q^{-1})(\cal{W}_0^2 + {\cal{W}_1}^2)-(q^2+q^{-2})(q-q^{-1})^2\epsilon_+\epsilon_-\gamma(q \cal{W}_1\cal{W}_0-q^{-1} \cal{W}_0\cal{W}_1)
\Big)\nonumber\\
&&\qquad \qquad-\Big( (\epsilon_+^2+\epsilon_-^2) \gamma^2 \frac{q-q^{-1}}{q^2+q^{-2}}-\epsilon_+^2\epsilon_-^2 \gamma^2  \frac{(q-q^{-1})^3}{\rho(q^2+q^{-2})}-\rho q^2 \frac{(\gamma^2-1)}{(q+q^{-1})^3(q^2+q^{-2})}\Big).
\ \nonumber 
\eeqa
The expressions for the elements $\tilde{\cal{G}}_{1},{\cal{W}}_{2},\tilde{\cal{G}}_{2}$ are obtained from ${\cal{G}}_{1},{\cal{W}}_{-1},{\cal{G}}_{2}$ by exchanging ${\cal W}_0\leftrightarrow{\cal W}_1$ in the above formulae. \vspace{1mm}

Actually, under certain assumptions it is possible to show that {\it all} descendants generators of the algebra ${\cal A}_q$ admit a unique explicit polynomial formulae in terms of the two fundamental generators $\cW_0,\cW_1$ \cite{BB1}. Roughly speaking, they take the form:
\beqa
{\cW}_{-k} &=& {\cal F}_{-k}({\cal W}_0,{\cal W}_1) , \qquad   {\cG}_{k+1} = {\cal F}_{k+1}({\cal W}_0,{\cal W}_1) , \label{poly}\\
{\cW}_{k+1} &=& {\cal F}_{-k}({\cal W}_1,{\cal W}_0) , \qquad   \tilde{\cG}_{k+1} = {\cal F}_{k+1}({\cal W}_1,{\cal W}_0) \qquad \mbox{for all} \quad k\in{\mathbb N},\nonumber
\eeqa
where $\{{\cal F}_{k}(X,Y),k\in{\mathbb Z}\}$ is a family of  two-variable polynomials in $X,Y$.  Note that the polynomial formulae that are obtained in \cite{BB1} can be independently derived using the connection between the algebra ${\cal A}_q$ and the reflection equation algebra \cite{BSh1}. Details  will be reported elsewhere.\vspace{1mm}

We now show how each realization of the four different types of $q-$Onsager algebras in terms of $U_q(\widehat{sl_2})$ Chevalley generators (\ref{realop}), (\ref{realopt}), (\ref{realopd}), (\ref{realopi})  can be recovered in a straightforward manner starting from the polynomial formulae (\ref{defel}) for the first few descendant generators. Define:
\beqa
\cW_0&=&\psi(\textsf W_0),\qquad{\cW}_1= \psi(\textsf W_1),\qquad \gamma = \psi(\Gamma), \label{realopbis}
\eeqa
with (\ref{realop}).  Observe that the polynomials (\ref{poly}) can be systematically expanded in terms of the Chevalley generators and parameters $k_\pm,\epsilon_\pm$. The explicit expressions for the first few examples (\ref{defel}), sufficient for our purpose, are reported in Appendix A. Then, we specialize case by case the boundary parameters into these expressions. According to the specialization chosen, the corresponding `reduced' descendant generators do produce the symmetry operators of each of the $q-$Onsager algebras exhibited in the previous Section. Explicitly, we obtain: \vspace{2mm}  

$\bullet$ {\bf Triangular boundary conditions  $k_-\neq 0$, $k_+=0$, $\epsilon_\pm\neq 0$}.
\beqa
{\cW}_0|_{k_+ \to 0}=\psi_t({\textsf T}_0),\qquad
{\cW}_1|_{k_+ \to 0}=\psi_t({\textsf T}_1),\qquad
\frac{\tilde\cG_1}{k_+}\Big|_{k_+ \to 0}=\psi_t(\tilde{\textsf P}_1).\nonumber
\eeqa

$\bullet$ {\bf Diagonal generic boundary conditions $k_\pm= 0,\epsilon_\pm\neq 0$}.
\beqa
\cW_0|_{k_\pm \to 0}=\psi_d({\textsf K}_0), \quad \cW_1|_{k_\pm \to 0}=\psi_d({\textsf K}_1),\quad
 \frac{\cG_1}{k_-}\Big|_{k_\pm \to 0}=\psi_d({\textsf Z}_1),\quad
\frac{\tilde\cG_1}{k_+}\Big|_{k_\pm \to 0}=\psi_d(\tilde{\textsf Z}_1).\nonumber
\eeqa

$\bullet$ {\bf Special diagonal boundary conditions  $k_\pm= 0,\epsilon_-=0,\epsilon_+=1$}.
\ben
&&{\textsf W}_1|_{k_\pm \to 0,\,\epsilon_+ \to 1, \epsilon_- \to 0}=0,\qquad \qquad \qquad \qquad \qquad \qquad {\textsf W}_0|_{k_\pm \to 0,\,\epsilon_+ \to 1, \epsilon_- \to 0}=\psi_i\left(\Gamma q^{-{\textsf h}}\right),\nonumber\\
&& \frac{\cG_1}{k_-}\Big|_{k_\pm \to 0, \,\epsilon_+ \to 1, \epsilon_- \to 0}=\psi_i\left((q^2-q^{-2})\Gamma q^{-1}{\textsf e}  q^{-{\textsf h}}\right), \qquad \qquad \frac{\tilde\cG_1}{k_+}\Big|_{k_\pm \to 0,\,\epsilon_+ \to 1, \epsilon_- \to 0} =\psi_i\left((q^2-q^{-2}) \Gamma  {\textsf f}\right),\nonumber\\
&& \cW_{2}|_{k_\pm \to 0,\,\epsilon_+ \to 1, \epsilon_- \to 0}=\psi_i\left(\frac{\Gamma}{q+q^{-1}}\left(q^{1+{\textsf h}} + q^{-1-{\textsf h}} +(q-q^{-1})^2{\textsf f}{\textsf e}\right)\right),\nonumber\\
&& \cW_{-1}|_{k_\pm \to 0,\,\epsilon_+ \to 1, \epsilon_- \to 0}=\psi_i\left(\frac{q-q^{-1}}{(q+q^{-1})^2}{\textsf X}q^{-{\textsf h}}\Gamma\right),\nonumber\\
&& \frac{\cG_2}{k_-}\Big|_{k_\pm \to 0, \,\epsilon_+ \to 1, \epsilon_- \to 0}= \psi_i\left( -q^{-2}\frac{(q-q^{-1})}{(q+q^{-1})}\Gamma{\textsf Y}q^{-{\textsf h}} +
q^{-1}\frac{(q-q^{-1})^2}{(q+q^{-1})}\Gamma{\textsf e}{\textsf X}q^{{\textsf h}}\right),\nonumber \\
&&   \frac{\tilde\cG_2}{k_+}\Big|_{k_\pm \to 0, \,\epsilon_+ \to 1, \epsilon_- \to 0}=
\psi_i\left(-q\frac{(q-q^{-1})}{(q+q^{-1})}\Gamma\tilde{\textsf Y} +
\frac{(q-q^{-1})^2}{(q+q^{-1})}{\textsf f}{\textsf X}\right).\nonumber
\een
\vspace{1mm}

Finally, as an alternative check of the analysis of the previous Section, we now show that according to the choice of boundary conditions, the corresponding above set of operators acting on the vector space (\ref{halfV}) commute with the associated Hamiltonian. For generic boundary conditions $k_\pm, \epsilon_\pm$, from (\ref{Hcom}) recall that the two operators $\cW_0,\cW_1$ acting on (\ref{halfV}) commute with the Hamiltonian. As a corollary of (\ref{Hcom}) and (\ref{poly}), for generic boundary conditions it implies that any descendant generator  is commuting with (\ref{Hsemi}):
\beqa
\big[ H_{\frac{1}{2}XXZ},{\cal F}_{k}(\cW_0^{(\infty)},\cW_1^{(\infty)})\big]=0, \qquad \big[ H_{\frac{1}{2}XXZ},{\cal F}_{k}(\cW_1^{(\infty)},\cW_0^{(\infty)})\big]=0 \qquad \mbox{for all} \quad k\in {\mathbb Z}.
\eeqa
Specializing the boundary parameters accordingly, it implies that the four different types of $q-$Onsager algebras are symmetry algebras of the Hamiltonian.\vspace{1mm}

Let us remark that the defining relations of the current algebra associated with the $q-$Onsager algebra are given in \cite{BSh1}. For the triangular, augmented and $U_q(gl_2)$ invariant $q-$Onsager algebras, the defining relations of the corresponding current algebras can be derived, respectively, by taking appropriate limits of the relations in $\cA_q$, given by \cite[Definition 3.1]{BSh1} and \cite[Proposition 3.1]{BSh1}.

\section{Concluding remarks}
It is instructive to consider the limit $q \to 1$ of the four different types of $q-$Onsager algebras described in Section 2. In particular, in this limit the $q$-Onsager, the augmented $q$-Onsager and the $U_q(gl_2)$ invariant $q$-Onsager algebras specialize to three different invariant fixed-point subalgebras of $U(\widehat{{sl_2}})$. For the  $q$-Onsager algebra with $q\rightarrow 1$, the corresponding automorphism is the Chevalley involution of $U(\widehat{{sl_2}})$, given by $\theta(e_i)=f_i$, $\theta(h_i)=-h_i$. For the augmented  $q$-Onsager algebra with $q\rightarrow 1$, one considers the composition of the Chevalley involution with the outer automorphism of $U(\widehat{{sl_2}})$. It reads:  $\theta_d(e_0)=f_1$, $\theta_d(e_1)=f_0$, $\theta_d(f_0)=e_1$, $\theta_d(f_1)=e_0$ and  $\theta_d(h_0)=-h_1$.  For the $U_q(gl_2)$ invariant $q$-Onsager algebra at $q\rightarrow 1$, one considers the composition of the Chevalley involution with the Lusztig's automorphism of $U(\widehat{{sl_2}})$ \cite{Lusz}. It reads: $\theta_i(e_0)=e_0$, $\theta_i(f_0)=f_0$, $\theta_i(h_0)=h_0$, $\theta_d(f_1)=[[e_1,e_0],e_0]/2$, $\theta_d(e_1)=[[f_1,f_0],f_0]/2$ and  $\theta_i(h_1)=-h_1-2 h_0$. From that point of view, three of the algebras introduced in Section 2 can be understood as a $q-$deformation of these invariant fixed-point subalgebras of $U(\widehat{{sl_2}})$ \cite{Kolb}. For the $q-$Onsager and augmented $q-$Onsager algebras, see  \cite{BC12} for details. Note that it is a simple exercice to apply the technique of \cite{BC12} to the $U_q(gl_2)$ invariant $q$-Onsager algebra. Besides these three algebras, let us point out that the triangular  $q$-Onsager algebra gives an example of coideal subalgebra that does not correspond to an invariant fixed-point subalgebra of $U_q(\widehat{{sl_2}})$ even at $q\rightarrow 1$.\vspace{1mm}

From the point of view of physics, whereas the existence of infinite Abelian symmetries in a lattice system - that are associated with infinitely many mutually commuting
conservation laws -  reduces the problem of degeneracies of the Hamiltonian's spectrum from infinite to finite, 
the existence of non-Abelian symmetries imply that common eigenspaces can be understood as irreducible modules of the symmetry 
algebra. For the thermodynamic limit of the open XXZ spin chain with Hamiltonian (\ref{Hsemi}) and certain boundary conditions it follows that the generators of the corresponding $q-$Onsager algebra discussed in Section 2 change eigenvectors of the Hamiltonian without changing the eigenvalues. Then, the solution of the model (spectrum, eigenstates, multiple integral representations of correlation functions and form factors) can be derived using  infinite dimensional ($q-$vertex operators) representations of the symmetry algebra. Clearly, the $q-$vertex operators for each symmetry algebra follow from the representation theory of $U_q(\widehat{sl_2})$. Note that the solution for special, diagonal and triangular boundary conditions is given in \cite{JKKKM,BB3,BKo1,BKo2}. For the Hamiltonian with generic boundary conditions, although $q-$vertex operators for the $q-$Onsager algebra are known \cite{BB3}, the solution remains an open problem.\vspace{1mm}

Finally, an interesting problem would be to extend the analysis here presented to integrable  models with higher rank symmetries of $q-$Onsager's type (see \cite{BB5}). The higher rank infinite dimensional algebra extending ${\cal A}_q$, yet unknown, could be a starting point for the identification of the symmetry algebras.

\vspace{0.3cm}

\noindent{\bf Acknowledgements:}  
P.B. thanks T. Kojima for discussions.
S.B. thanks V. Regelskis for discussions and also LMPT for hospitality, where part of this work has been done. S.B. is supported by a public grant as part of the
"investissement d'avenir" project, reference ANR-11-LABX-0056-LMH,
LabEx LMH. P.B. is supported by C.N.R.S.

\vspace{0.4cm}

\begin{center} {\underline{APPENDIX A:} Descendant generators in the Chevalley-type presentation of $U_q(\widehat{sl_2})$}\end{center}
\vspace{1mm}
\ben
\label{G1sc}
\frac{{\cG}_{1}}{k_-}&=&(q^2-q^{-2})\Big(\epsilon_- q^c f_1+\epsilon_+ q^{-1} e_0 q^{h_1}+k_-f_1q^{h_1}e_0\Big)\\
&&+ \ k_+\big(q e_1e_0-q^{-1}e_0e_1+(q f_0f_1-q^{-1}f_1f_0)q^c \big), \nonumber
\een
\ben
\label{W-1sc}
{\cW}_{-1}&=&\frac{\epsilon_-}{q+q^{-1}} \Big(q^{1+h_1}+q^{-1-h_1}+(q-q^{-1})^2 f_1e_1\Big)q^c\\
&&+\ \epsilon_+ \frac{(q-q^{-1})}{(q+q^{-1})^2}\Big(q e_1e_0-q^{-1}e_0e_1+(q f_0f_1-q^{-1} f_1f_0)q^c\Big)q^{h_1}\nonumber\\
&&+\ \frac{k_+}{(q+q^{-1})^2}\Big((q-q^{-1})(q f_0f_1-q^{-1}f_1f_0)q^c e_1+q^{-2}(q+q^{-1}) f_0q^{h_0}\nonumber\\
&&Ê\qquadÊ\qquad \qquad \qquad-(e_1)^2e_0+(q^2+q^{-2})e_1e_0e_1-e_0(e_1)^2\Big)\nonumber\\
&&+\ \frac{k_-}{(q+q^{-1})^2}\Big(q^{-1} (q+q^{-1})e_0+q^{-1}(q-q^{-1})f_1q^{h_1}(q e_1e_0-q^{-1}e_0e_1)\nonumber\\
&&\qquad\qquad \qquad \qquad -q^{c-1}\big((f_1)^2f_0-(q^2+q^{-2})f_1f_0f_1+f_0(f_1)^2\big)\Big),\nonumber
\een

\ben
\label{G2sc}
\frac{{\cG}_{2}}{k_-}&=& \epsilon_-\frac{q-q^{-1}}{q+q^{-1}} \Big((q+q^{-1})q^c e_0 q^{h_1}+\frac{q^2-q^{-2}}{q+q^{-1}} q^cf_1(q e_0e_1-q^{-1} e_1e_0) \\
&& \qquad\qquad\qquad \qquad -q^{2c}\big((f_1)^2f_0-(q^2+q^{-2})f_1f_0f_1+f_0(f_1)^2\big)\Big)\nonumber\\
&& + \  \epsilon_+\frac{q-q^{-1}}{q+q^{-1}} \Big((q+q^{-1})q^{2c+1} f_1+\frac{q^2-q^{-2}}{q+q^{-1}} q^{c-1}(q f_0f_1-q^{-1} f_1f_0)e_0 q^{h_1}\nonumber \\
&& \qquad\qquad\qquad \qquad -q^{-1}\big((e_0)^2e_1-(q^2+q^{-2})e_0e_1e_0+e_1(e_0)^2\big)\Big)+O(k_+)+O(k_-).\nonumber
\eeqa 

Note that for simplicity, the terms of order $k_+$ and $k_-$ in $\frac{{\cG}_{2}}{k_-}$ are not explicitly written, as they do not contribute in the limit $k_\pm=0$. Also, the descendant generators $\frac{{\tilde \cG}_{1}}{k_+}$ , ${\cW}_{2}$  and $\frac{{\tilde \cG}_{2}}{k_+}$ can be derived in terms of the generators of the Chevalley-type presentation using the map $x_0 \to x_1$, $x_1 \to x_0$, with $x \in \{ h,e,f\}$, $k_\pm \to k_\mp$ and $\epsilon_\pm \to \epsilon_\mp$ on the expressions (\ref{G1sc}), (\ref{W-1sc}) and (\ref{G2sc}) respectively.

\vspace{0.5cm}


\begin{thebibliography}{DJKM}




\bibitem[B04]{Bas1}
P. Baseilhac, {\it Deformed Dolan-Grady relations in quantum integrable models}, Nucl.Phys. {\bf B 709} (2005) 491-521, {\tt arXiv:hep-th/0404149}.
%
\bibitem[B05]{Bas2}
P. Baseilhac, {\it An integrable structure related with tridiagonal algebras}, Nucl.Phys. {\bf B 705} (2005) 605-619, {\tt arXiv:math-ph/0408025}. 
%

\bibitem[BB09]{BB5}
P. Baseilhac and S. Belliard, {\it Generalized q-Onsager algebras and boundary affine Toda field theories}, Lett. Math. Phys. {\bf 93} (2010) 213-228, {\tt arXiv:0906.1215}.


%
\bibitem[BB10]{BB1}
P. Baseilhac and S. Belliard, {\it A note on the $O_q(\widehat{sl_2})$ algebra}, {\tt arXiv:1012.5261}.


\bibitem[BB13]{BB3}
P. Baseilhac and S. Belliard, {\it The half-infinite XXZ chain in Onsager's approach}, Nucl. Phys. {\bf B 873} 550-583 (2013), {\tt 
arXiv:1211.6304}.


\bibitem[BK05]{BK1}
P. Baseilhac and K. Koizumi, {\it A new (in)finite dimensional algebra for quantum integrable models}, Nucl. Phys. {\bf B 720} (2005) 325-347, {\tt arXiv:math-ph/0503036}.
%

%
\bibitem[BK07]{BK2}
P. Baseilhac and K. Koizumi, {\it A deformed analogue of Onsager's symmetry in the XXZ open spin chain}, J.Stat.Mech. {\bf 0510} (2005) P005, {\tt arXiv:hep-th/0507053};\\
(-----), {\it Exact spectrum of the XXZ open spin chain from the q-Onsager algebra representation theory}, J. Stat. Mech.  (2007) P09006, {\tt arXiv:hep-th/0703106}.


\bibitem[BKo13]{BKo1}
P. Baseilhac and T. Kojima, {\it Correlation functions of the half-infinite XXZ spin chain with a triangular boundary}, Nucl. Phys. {\bf B 880} (2014) 378–413, {\tt  arXiv:1309.7785}.
    
\bibitem[BKo14]{BKo2}
P. Baseilhac and T. Kojima, {\it Form factors of the half-infinite XXZ spin chain with a triangular boundary}, J. Stat. Mech. (2014) P09004, {\tt   arXiv:1404.0491}.
    

%
\bibitem[BS09]{BSh1}
P. Baseilhac and K. Shigechi, {\it A new current algebra and the reflection equation}, Lett. Math. Phys. {\bf 92} (2010) 47-65, {\tt arXiv:0906.1482}.
%

\bibitem[BC12]{BC12}
S. Belliard and N. Cramp{\'e}, {\it Coideal algebras from twisted Manin triples}, Journal of Geometry and Physics {\bf 62} (2012), pp. 2009-2023

%
\bibitem[D90]{Davies}
B. Davies, {\it Onsager's algebra and superintegrability}, J. Phys. {\bf A 23} (1990) 2245-2261;\\
B. Davies, {\it Onsager's algebra and the Dolan-Grady condition in the non-self-dual case}, J. Math. Phys. {\bf 32} (1991) 2945-2950.
%


%
\bibitem[DFJMN92]{vertex}
M. Jimbo, K. Miki, T. Miwa and  A. Nakayashiki, {\it Correlation functions of the XXZ model for $\Delta<-1$ }, Phys Lett. A 168 (1992) 256, {\tt arXiv:hep-th/9205055};\\
B. Davies, O. Foda, M. Jimbo, T. Miwa and A. Nakayashiki, {\it Diagonalization of the XXZ Hamiltonian by vertex operators }, Commun. Math. Phys {\bf 151} (1993) 89, {\tt arXiv:hep-th/9204064};\\
M. Jimbo, T. Miwa, {\it qKZ equation with $|q|=1$ and correlation functions of the XXZ model in the gapless regime}, J. Phys. {\bf A 29} (1996) 2923, {\tt arXiv:hep-th/9601135}.
%


%
\bibitem[DG82]{DG}
L. Dolan and M. Grady, {\it Conserved charges from self-duality}, Phys. Rev. {\bf D 25} (1982) 1587.
%

%
\bibitem[FM92]{FM}
O. Foda and T. Miwa, {\it Corner transfer matrices and quantum affine algebras}, Int. J. Mod. Phys. {\bf A 7} Suppl. 1A (1992), 279-302.  {\tt arXiv:hep-th/9204068v1}.
%



%
\bibitem[IIJMNT92]{Idz}
M. Idzumi, K. Iohara, M. Jimbo, T. Miwa, T. Nakashima and T. Tokihiro, {\it Quantum affine symmetry in vertex models}, Int. Jour. Mod. Phys. {\bf 8} (1993) 8, 1479-1511; {\tt arXiv:hep-th/9208066v1}.


%
\bibitem[IT09]{IT}
T. Ito and P. Terwilliger, {\it The augmented tridiagonal algebra}, Kyushu Journal of Mathematics, {\bf 64} (2010) No. 1, 81-144, {\tt arXiv:0904.2889v1}. 


%
\bibitem[J92]{Jim0}
M. Jimbo, {\it Quantum group symmetry and lattice correlation functions}, Chinese Journal of Physics {\bf 30} (1992) 7, 973-985.
%
\bibitem[JKKKMW95]{JKKKM}
M. Jimbo, R. Kedem, T. Kojima, H. Konno and T. Miwa, {\it XXZ chain with a boundary}, Nucl. Phys. B 441 (1995) 437-470;\\ 
M. Jimbo, R. Kedem, H. Konno, T. Miwa and R. Weston, {\it Difference Equations in Spin Chains with a Boundary}, Nucl. Phys. B 448 (1995) 429-456. 
%

\bibitem[Koj10]{Koj}
T. Kojima, {\it Diagonalization of transfer matrix of supersymmetry $U_q(sl(m+1|n+1))$ chain with a boundary}, 
{\tt arXiv:1211.2912}, J. Math. Phys. {\bf 54} (2013) 043507, \\
(---), {\it A remark on ground state of boundary Izergin-Korepin model},  Int. J. Mod. Phys. {\bf A 26} (2011) 1973, {\tt arXiv:1101.4078},\\
(---), {\it Diagonalization of infinite transfer matrix of boundary $U_{q,p}(A^{(1)}_{N-1})$ face model}, 
J. Math. Phys. {\bf 52} (2011) 013501.  
{\tt arXiv:1007.3795}. 


\bibitem[Kol12]{Kolb}
S. Kolb,  {\it Quantum symmetric Kac-Moody pairs}, Adv. Math. {\bf 267} (2014), 395-469, {\tt arXiv:1207.6036v1}.


\bibitem[Koy93]{Koy}
 Y. Koyama, {\it Staggered Polarization of Vertex Models with $U_q(\widehat{sl(n)})$-Symmetry}, Commun. Math. Phys. {\bf 164} (1994) 277-292, {\tt arXiv:hep-th/9307197}. 

\bibitem[L93]{Lusz}
G. Lusztig, {\it Introduction to Quantum Groups}, Birkhauser (1993).
    

\bibitem[Ons44]{Ons44}
L. Onsager, {\it Crystal Statistics. I. A Two-Dimensional Model with an Order-Disorder Transition}, Phys. Rev. {\bf 65} (1944) 117-149.
%


\bibitem[PS90]{PS}
V. Pasquier and H. Saleur, {\it Common structures between finite systems and conformal field theories through quantum groups}, Nucl. Phys. {\bf B 330} (1990) 523-556.

%
\bibitem[R12]{Vidas}
V. Regelskis, {\it Reflection algebras for $SL(2)$ and $GL(1|1)$}, {\tt arXiv:1206.6498}.


\bibitem[T03]{Ter03}
P. Terwilliger,{\it Two relations that generalize the $q-$Serre
relations and the Dolan-Grady relations}, Proceedings of the Nagoya 1999 International workshop on physics and combinatorics. Editors A. N. Kirillov, A. Tsuchiya, H. Umemura. 377-398, {\tt math.QA/0307016}.
%

\end{thebibliography}
\end{document}